\documentclass[12pt]{iopart}
\usepackage{epsfig}
\begin{document}

\title[Latest Results from PHOBOS]{Latest Results from PHOBOS}

\author{D~J~Hofman$^6$ for the PHOBOS Collaboration\\
\vspace{0.2in}
%
% Authors for preliminary data shown at QM06 which included:
% RUN2005: CuCu @ 22, 63 & 200 GeV and pp @ 400 GeV
% RUN2004: AuAu @ 63 & 200 GeV and pp @ 200 GeV
%
% Last edited 23-Jan-2007 by George Stephans\\  \vspace{0.2in}
%
B.Alver$^4$,
B.B.Back$^1$,
M.D.Baker$^2$,
M.Ballintijn$^4$,
D.S.Barton$^2$,
R.R.Betts$^6$,
A.A.Bickley$^7$,
R.Bindel$^7$,
W.Busza$^4$,
A.Carroll$^2$,
Z.Chai$^2$,
V.Chetluru$^6$,
M.P.Decowski$^4$,
E.Garc\'{\i}a$^6$,
N.George$^2$,
T.Gburek$^3$,
K.Gulbrandsen$^4$,
C.Halliwell$^6$,
J.Hamblen$^8$,
I.Harnarine$^6$,
M.Hauer$^2$,
C.Henderson$^4$,
D.J.Hofman$^6$,
R.S.Hollis$^6$,
R.Ho\l y\'{n}ski$^3$,
B.Holzman$^2$,
A.Iordanova$^6$,
E.Johnson$^8$,
J.L.Kane$^4$,
N.Khan$^8$,
P.Kulinich$^4$,
C.M.Kuo$^5$,
W.Li$^4$,
W.T.Lin$^5$,
C.Loizides$^4$,
S.Manly$^8$,
A.C.Mignerey$^7$,
R.Nouicer$^2$,
A.Olszewski$^3$,
R.Pak$^2$,
C.Reed$^4$,
E.Richardson$^7$,
C.Roland$^4$,
G.Roland$^4$,
J.Sagerer$^6$,
H.Seals$^2$,
I.Sedykh$^2$,
C.E.Smith$^6$,
M.A.Stankiewicz$^2$,
P.Steinberg$^2$,
G.S.F.Stephans$^4$,
A.Sukhanov$^2$,
A.Szostak$^2$,
M.B.Tonjes$^7$,
A.Trzupek$^3$,
C.Vale$^4$,
G.J.van~Nieuwenhuizen$^4$,
S.S.Vaurynovich$^4$,
R.Verdier$^4$,
G.I.Veres$^4$,
P.Walters$^8$,
E.Wenger$^4$,
D.Willhelm$^7$,
F.L.H.Wolfs$^8$,
B.Wosiek$^3$,
K.Wo\'{z}niak$^3$,
S.Wyngaardt$^2$,
B.Wys\l ouch$^4$}

%
% Note that this is the full form of the addresses, for conference proceedings,
% you can use the reduced one that follows
%
% $^1$~Physics Division, Argonne National Laboratory, Argonne, IL 60439-4843,
% USA\\
% $^2$~Chemistry and C-A Departments, Brookhaven National Laboratory, Upton, NY
% 11973-5000, USA\\
% $^3$~Institute of Nuclear Physics, Krak\'{o}w, Poland\\
% $^4$~Laboratory for Nuclear Science, Massachusetts Institute of Technology,
% Cambridge, MA 02139-4307, USA\\
% $^5$~Department of Physics, National Central University, Chung-Li, Taiwan\\
% $^6$~Department of Physics, University of Illinois at Chicago, Chicago, IL
% 60607-7059, USA\\
% $^7$~Department of Chemistry, University of Maryland, College Park, MD 20742,
% USA\\
% $^8$~Department of Physics and Astronomy, University of Rochester, Rochester,
% NY 14627, USA\\
%
%
\address{
$^1$~Argonne National Laboratory, Argonne, IL 60439-4843, USA\\
$^2$~Brookhaven National Laboratory, Upton, NY 11973-5000, USA\\
$^3$~Institute of Nuclear Physics PAN, Krak\'{o}w, Poland\\
$^4$~Massachusetts Institute of Technology, Cambridge, MA 02139-4307, USA\\
$^5$~National Central University, Chung-Li, Taiwan\\
$^6$~University of Illinois at Chicago, Chicago, IL 60607-7059, USA\\
$^7$~University of Maryland, College Park, MD 20742, USA\\
$^8$~University of Rochester, Rochester, NY 14627, USA}

\ead{hofman@uic.edu}

\begin{abstract}
This manuscript contains a summary of the latest physics results 
from PHOBOS, as reported at Quark Matter 2006.  Highlights
include the first measurement from PHOBOS of dynamical elliptic 
flow fluctuations as well as an explanation of their possible 
origin, two-particle correlations, identified particle ratios, 
identified particle spectra and the latest results in global charged 
particle production.
\end{abstract}


%Uncomment for Submitted to journal title message
\submitto{\jpg}

%Comment out if separate title page not required
\maketitle

\normalsize

%\section{Introduction}

Over the last year, PHOBOS \cite{PHOBOS:nim} 
has continued to analyze the large dataset 
obtained from the first five runs (2000-2005) of the Relativistic 
Heavy Ion Collider (RHIC) at Brookhaven National Laboratory.  
To a large extent, one of the
primary goals of PHOBOS to obtain a broad survey of the global properites 
of charged particle production is nearing completion.  We report
two new results in this area, the first for the lowest energy Cu+Cu 
collisions, $\sqrt{s_{_{\rm NN}}}$ = 22.4 GeV, and the second for the
peripheral yields of midrapidity charged particle emission in Au+Au.
In both cases the new data has followed the striking scaling rules
reported in the past by PHOBOS \cite{Baker:qm02,PHOBOS:wp,Roland:qm05}, 
in particular the energy-independence of particle production
when viewed in the rest-frame of one of the 
colliding nuclei \cite{PHOBOS:els} and the factorization of 
the energy and centrality dependence of midrapidity charged 
particle yields \cite{PHOBOS:midrap-e-c-factor,PHOBOS:pt-e-c-factor}.
We also report final results for identified hadron transverse
momentum spectra down to very low $p_T$ in Au+Au collisions 
at $\sqrt{s_{_{\rm NN}}}$ = 62.4 GeV as well as new results for 
antiparticle to particle ratios in Cu+Cu 
collisions \cite{Veres:qm06}.
The measurement of both elliptic and directed flow
of charged particles over a uniquely large range in pseudorapidity
($-5.4<\eta<5.4$) has been an 
area of active analysis in PHOBOS from the very first Au+Au 
collisions \cite{PHOBOS:v2-130GeVAuAu,PHOBOS:v1-AuAu}.
The large elliptic flow signals ($v_2$) measured near midrapidity at full
RHIC energies, which are in agreement with hydrodynamical model
calculations of a relativistic hydrodynamic fluid, 
have provided strong credence for our current view that we are producing
a strongly interacting state of matter that reaches equilibration
early in the collision process.  
The comparison between Cu+Cu and Au+Au of $v_2(\eta)$ and mid-rapidity
$v_2(p_T)$ \cite{Nouicer:qm06}
further strengthens our understanding of the PHOBOS result
reported at Quark Matter 2005 \cite{Manly:qm05}: 
the $v_2$ results for both systems
can be consistently unified when scaled by the 
``participant eccentricity'',
$\langle\varepsilon_{part}\rangle$, where $\varepsilon_{part}$ is
defined event-by-event as the initial overlap eccentricity measured in
the rotated ``participant'' frame where that eccentricity is maximized.
Furthermore, the importance of the initial event-by-event interaction
points of the participant nucleons as defined by 
$\langle\varepsilon_{part}\rangle$, comes into
clear focus when the fluctuations of this quantity are calculated and
compared to the newly measured $v_2$ dynamical fluctuations in 
Au+Au \cite{Loizides:qm06,Alver:qm06}. 
On the two particle correlations front, we present new data showing
correlations between particles in p+p and Cu+Cu data at 
$\sqrt{s_{_{\rm NN}}}$ = 200 GeV over a uniquely large 
region of $\Delta\eta$ and $\Delta\phi$ \cite{Li:qm06}.  
We also present the first physics results of a study
of the centrality dependence of the short-range pseudorapidity 
correlations as measured by the effective cluster size, $K_{\rm eff}$.
Studies of cluster production indicate 
that the isotropic cluster model produces a cluster width in 
pseudorapidity that is larger than found in the resonance cascade 
model.  PHOBOS also reported on the latest results
of two ongoing studies of multiplicity fluctuations in Au+Au collisions,
in particular the search for enhanced $dN_{\rm ch}/d\eta$ fluctuations and
the foward-backward multiplicity fluctuations in the context of 
two models of cluster production \cite{Wozniak:qm06}.  The current 
upper limit for the fraction of events with large deviations in the
$dN_{\rm ch}/d\eta$ distribution of 200 GeV Au+Au collisions 
is $10^{-5}$.  
In conclusion, we will 
give a brief overview of the future direction of the PHOBOS 
research program.

\begin{figure}[t]
\centering
\includegraphics[width=0.55\textwidth]{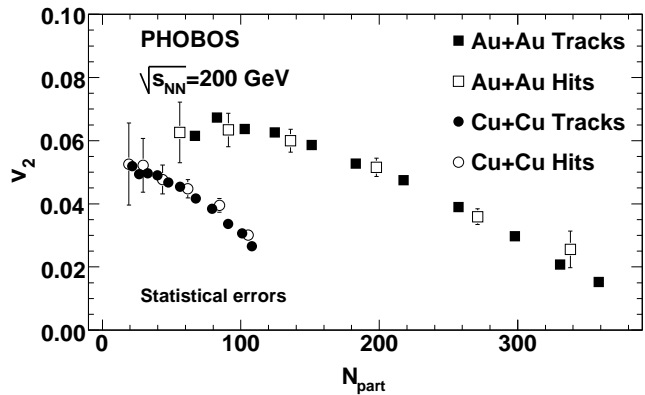}
\caption{Magnitude of the average elliptic flow coefficient, $v_2$, 
at midrapidity as a function of centrality ($N_{\rm part}$) for Cu+Cu 
and Au+Au collisions at $\sqrt{s_{_{\rm NN}}}$ = 200 GeV.  
\label{fig:1_v2}}
\end{figure}

One of the primary motivations of colliding Cu+Cu and Au+Au 
nuclei at RHIC was to enable a detailed study of the effect of 
system size on all measurable physics observables.  The study
of elliptic flow, with its sensitivity to the collectivity of
the produced matter at very early times of the collision, is 
a physics measurement of particular interest. The result of this
analysis and comparison for $\sqrt{s_{_{\rm NN}}}$ = 200 GeV is shown
in \Fref{fig:1_v2} for the PHOBOS hit-based and track-based
elliptic flow analysis, as a function of centrality 
defined by the number of participants, $N_{\rm part}$. Two 
important features are immediately evident.  First, the magnitude
of flow in the smaller Cu+Cu system is large and qualitatively 
follows a similar trend with centrality as seen in the larger Au+Au
system. Second, even for the most central collisions in Cu+Cu,
the magnitude of $v_2$ is substantial, and exceeds that seen in
central Au+Au.

\begin{figure}[ht]
\centering
\includegraphics[width=0.45\textwidth]{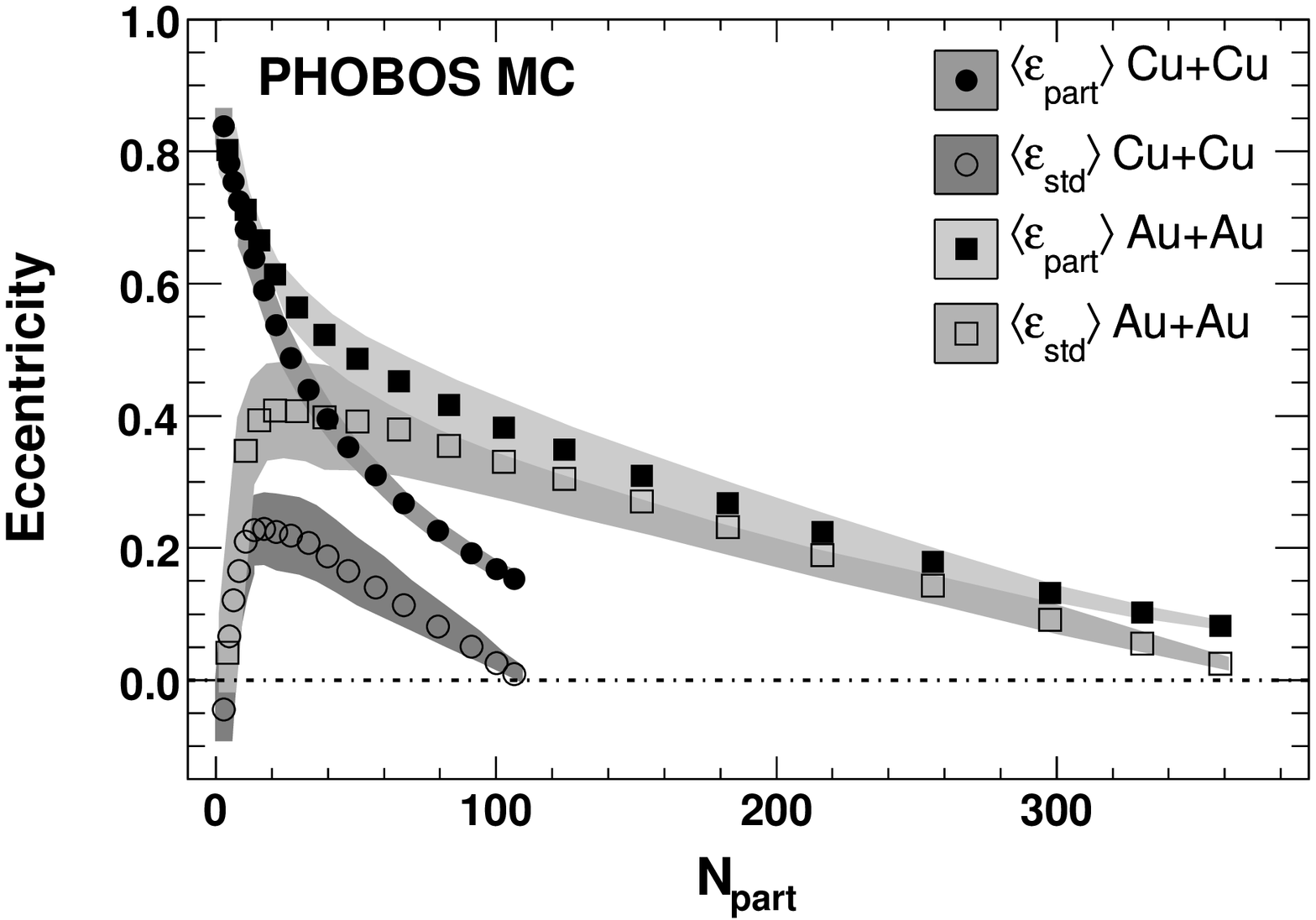}
\vspace{2pt}
%\hfill
\includegraphics[width=0.45\textwidth]{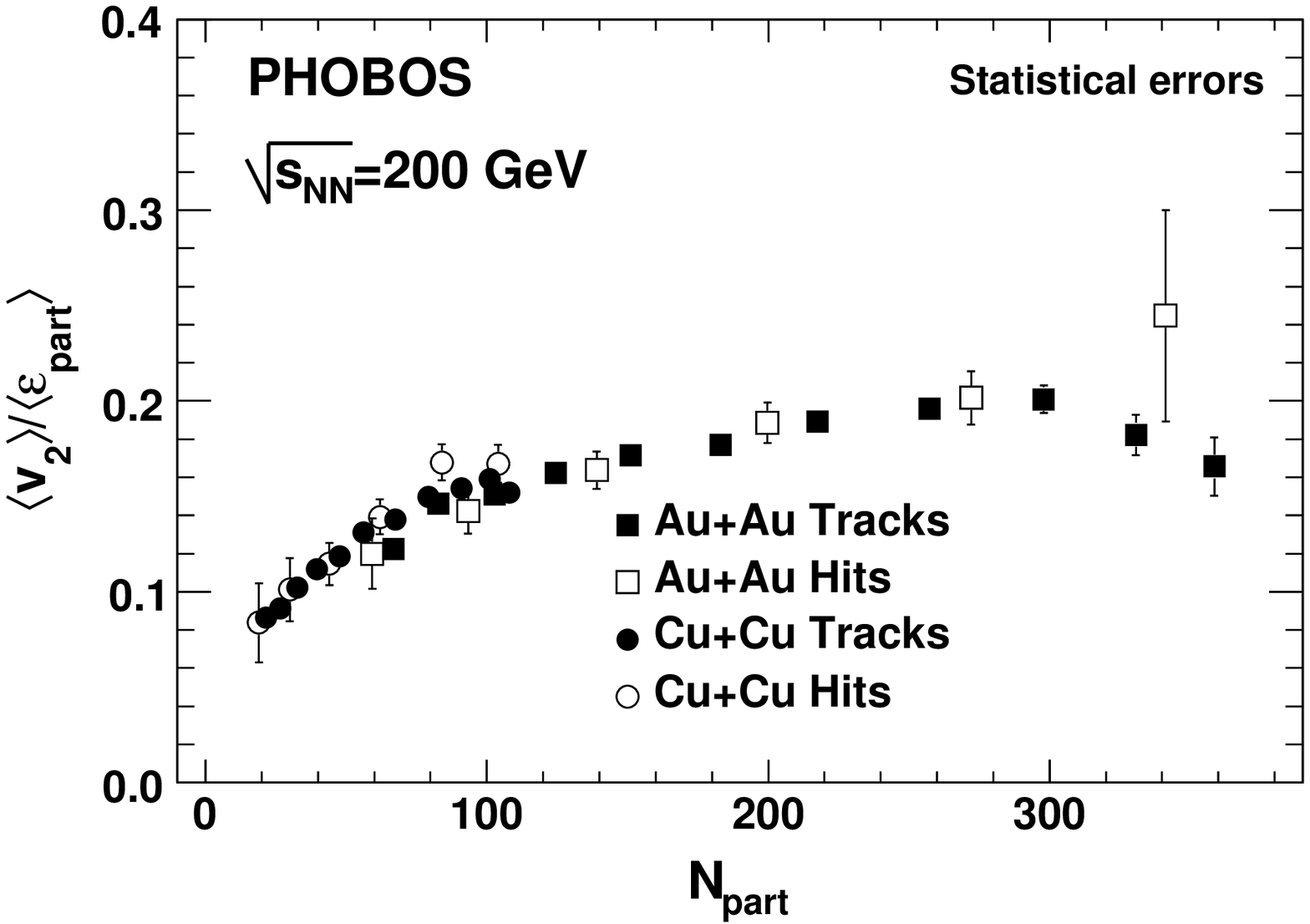}
\caption{Left: Two different
calculations for the average eccentricity for Cu+Cu and Au+Au
collisions at $\sqrt{s_{_{\rm NN}}}$ = 200 GeV, see text for details.
Right: The measured midrapidity $v_2$ divided 
by the participant eccentricity.
\label{fig:2_v2-epart}}
\end{figure}

In the most intuitive picture, our understanding of elliptic flow is
that the magnitude of the final azimuthal angular distribution 
of produced particles relative to the reaction plane is a consequence
primarily of the initial overlap eccentricity of the colliding nuclei. 
If this picture is correct, the elliptic flow results for both Cu+Cu
and Au+Au should be compatible if each
is scaled by the proper eccentricity.  We have proposed that
the most relevant eccentricity is not the one calculated relative to the
initial impact parameter vector, which we will call the `standard'
calculation $\langle\epsilon_{\rm std}\rangle$, but instead 
the eccentricity calculated event-by-event after rotating into the
frame of reference that maximizes the eccentricity defined by the 
participant nucleon interaction points, 
$\langle\epsilon_{\rm part}\rangle$ \cite{Manly:qm05,PHOBOS:v2-epart}.  
The difference between these calculations is shown 
in \Fref{fig:2_v2-epart}, and deviations are clearly evident
for smaller number of participants in Au+Au and all centralities
of Cu+Cu collisions, a result that illustrates the importance
of finite-number fluctuations of the participant interaction points.
This result is robust to the details of the Glauber Monte Carlo simulation,
as indicated by the bands which show the 90\% C.L. systematic errors.
When the $v_2$ data of \Fref{fig:1_v2} is scaled by 
$\langle\epsilon_{\rm part}\rangle$, the two very different systems
are unified on a single curve as shown in the right-hand side of
\Fref{fig:2_v2-epart}.

\begin{figure}[th]
\centering
\includegraphics[width=0.45\textwidth]{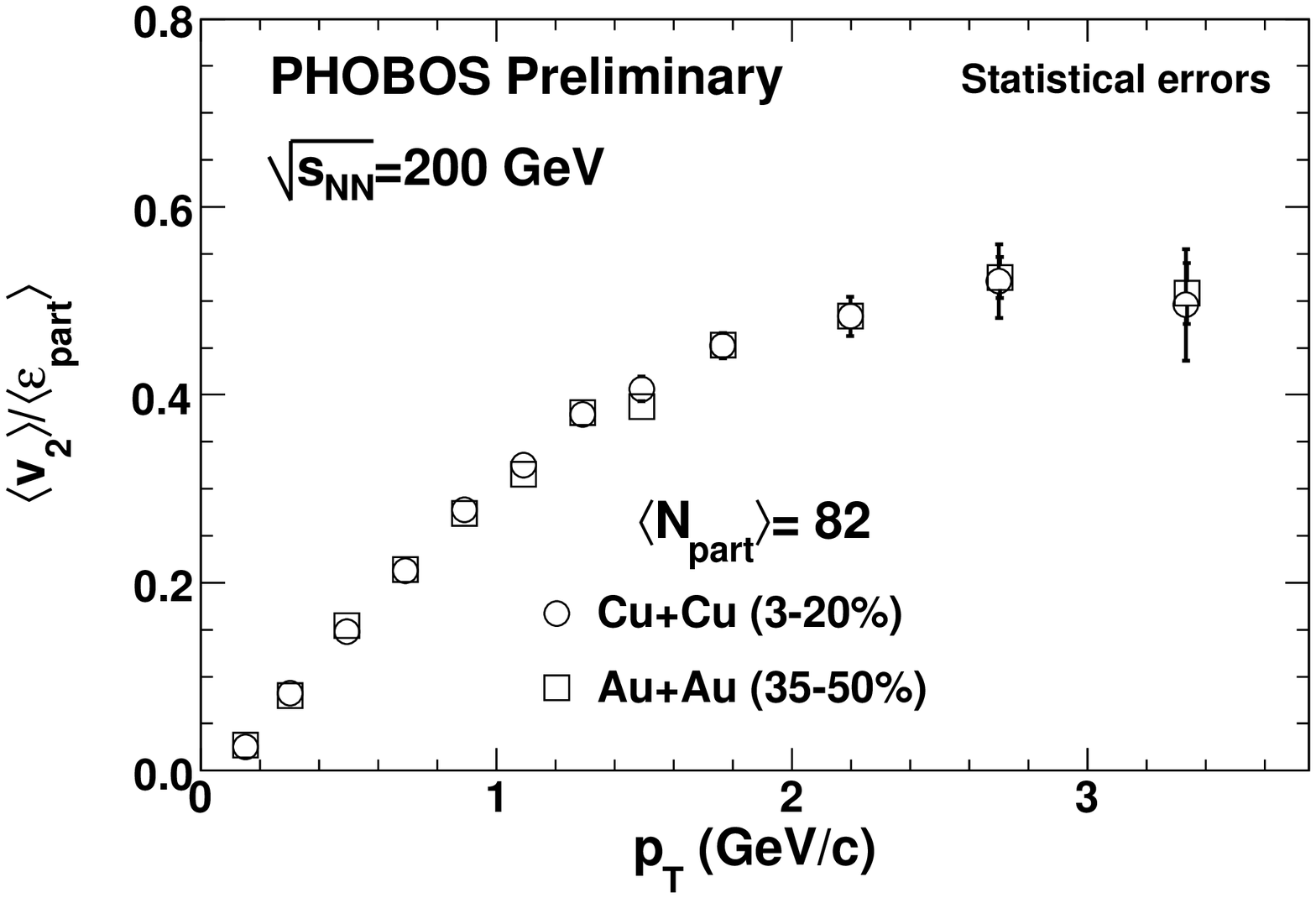}
\vspace{2pt}
%\hfill
\includegraphics[width=0.45\textwidth]{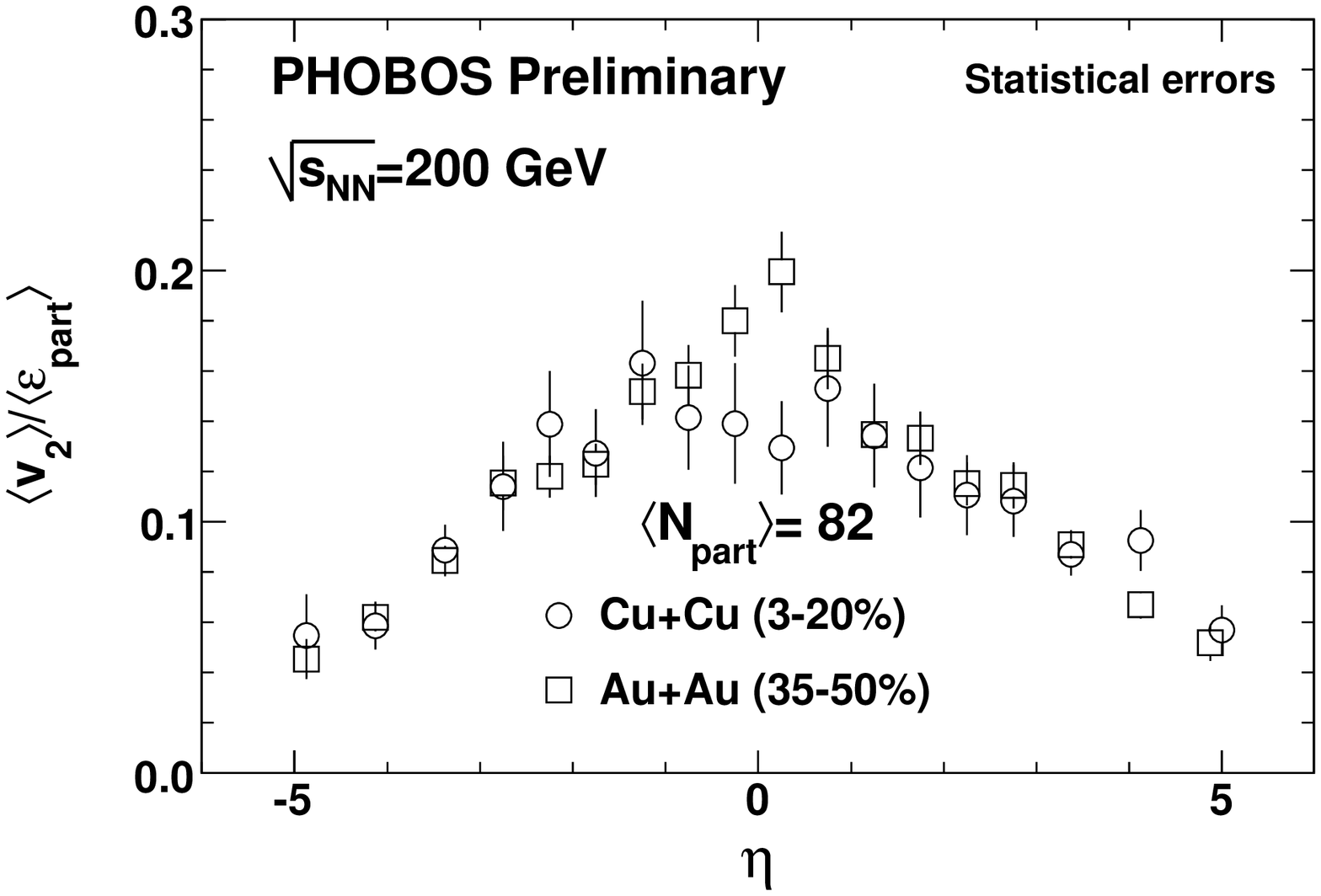}
\caption{Eccentricity scaled elliptic flow results 
in Cu+Cu and Au+Au collisions at matched $\langle N_{\rm part} \rangle$. 
Results are shown for $v_2/\langle\epsilon_{\rm part}\rangle$ 
versus $p_T$ (left) and $\eta$ (right).
\label{fig:3_v2_pt-eta}}
\end{figure}

The unification of the elliptic flow results in Cu+Cu and Au+Au 
collisions when scaled by the participant eccentricity holds
not only for the average value of $v_2$ at midrapidity, but also as a 
function of transverse momentum and pseudorapidity.  This new
result is shown in \Fref{fig:3_v2_pt-eta}, where
for the same number of participants in both systems we
find a consistent result out to 3.5 GeV/c in $p_T$ and across $\pm 5$ 
units of $\eta$.

\begin{figure}[t]
\centering
\includegraphics[width=0.45\textwidth]{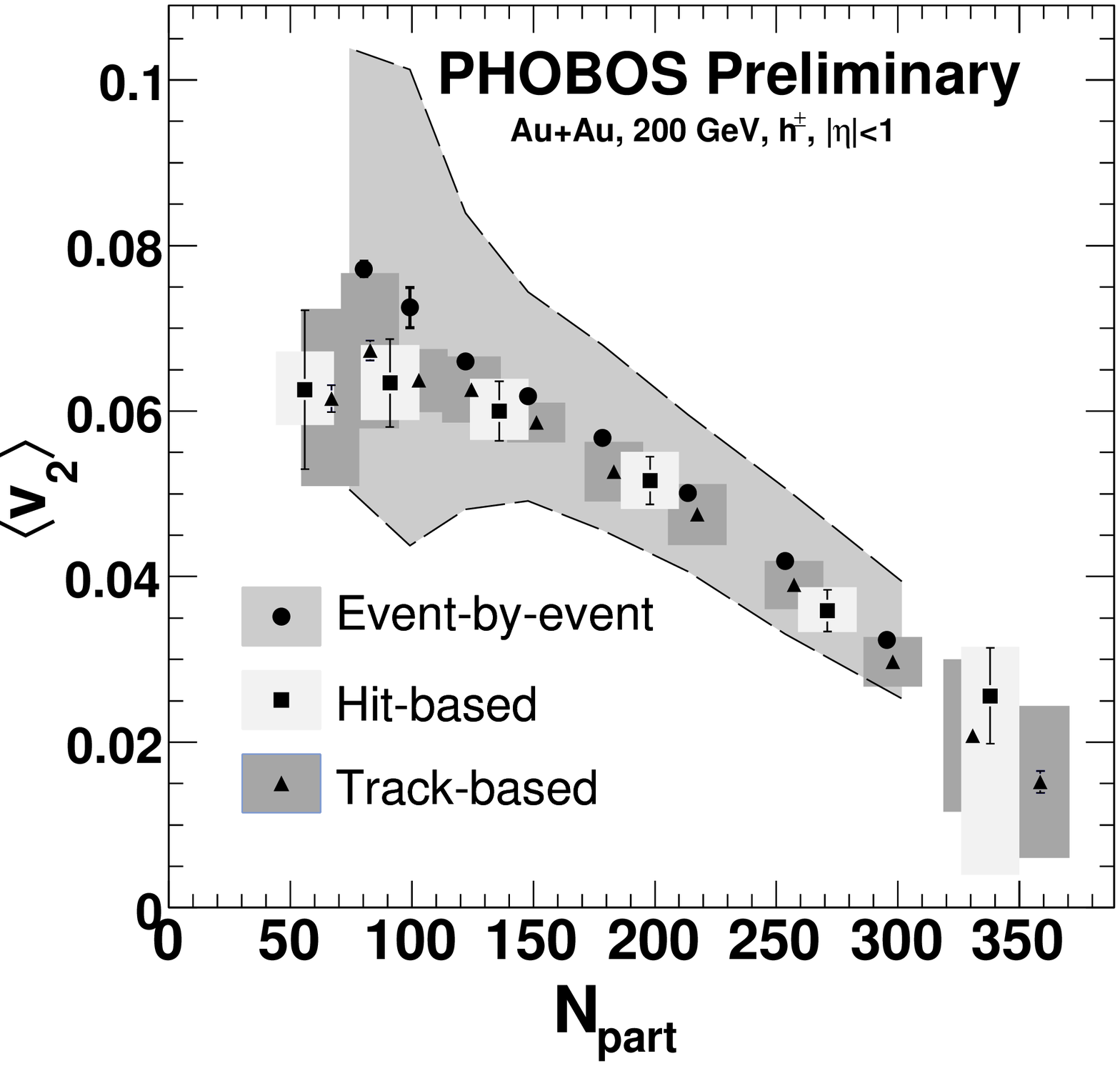}
\vspace{2pt}
%\hfill
\includegraphics[width=0.45\textwidth]{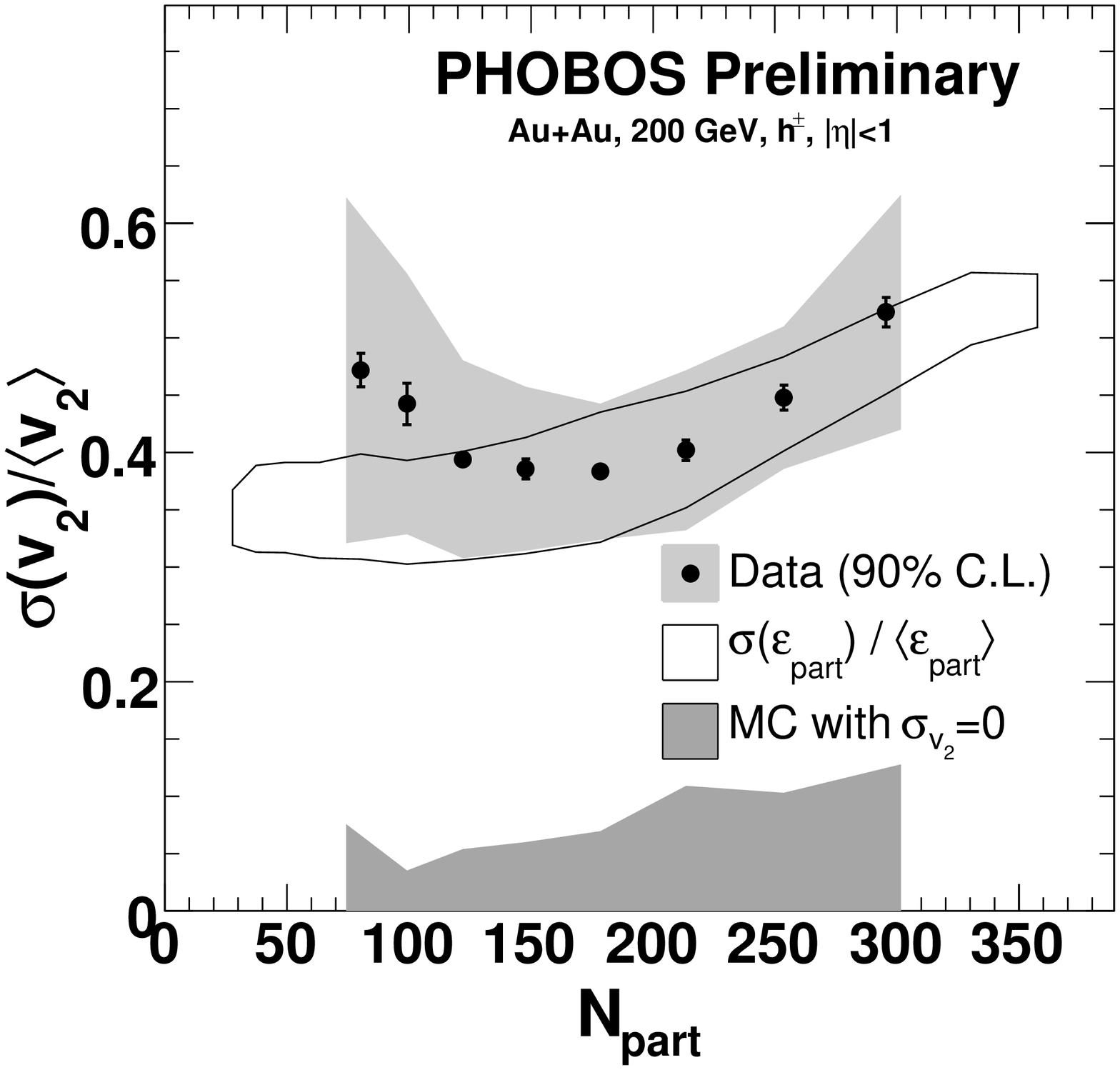}
\caption{Left: Event-by-event measurement
of $\langle v_2 \rangle$ compared with hit and track based PHOBOS results.
Right: Measured dynamical fluctuations in elliptic flow 
($\sigma(v_2)/\langle v_2 \rangle$), participant eccentricity
fluctuations calculated in the PHOBOS participant eccentricity model,
and the sensitivity of the measurement in the case of no dynamical
fluctuations.
\label{fig:4_v2_ebye}}
\end{figure}

The apparent relevance of the participant eccentricity model in unifying
the average elliptic flow results for Cu+Cu and Au+Au collisions leads
naturally to consideration of the dynamical fluctuations of both
the participant eccentricy itself as well as in the elliptic flow
signal from data.  Simulations of the expected dynamical fluctuations 
in participant eccentricity as a function of $N_{\rm part}$
were performed using the PHOBOS Monte Carlo Glauber based
participant eccentricity model, and they predict
large dynamical fluctuations,
$\sigma(\epsilon_{\rm part})/\langle\epsilon_{\rm part}\rangle$ of the
order of 0.4 in Au+Au collisions at $\sqrt{s_{_{\rm NN}}} = 200$ GeV.
There are several different approaches one could develop
to measure dynamical elliptic flow fluctuations, and PHOBOS has recently
created a new method that is based on a direct measure of $v_2$ on 
an event-by-event basis using a maximum likelihood fit that utilizes
the unique large pseudorapidity coverage of the PHOBOS 
detector \cite{Alver:qm06,Alver:ebye_v2_method}.  The strength of this
approach lies in the fact that this analysis removes the 
effects of statistical fluctuations and multiplicity dependence 
by applying a detailed model of the detector response 
that enables both a measurement 
of the average $v_2$ 
on an event-by-event basis as well as a measure of the dynamical
fluctuations in $v_2$.
The experimental results for both the average $\langle v_2 \rangle$ and the
measured dynamical fluctuations, which we quantify using the ratio
$\sigma(v_2)/\langle v_2 \rangle$, obtained in this new analysis
are given in \Fref{fig:4_v2_ebye}
for Au+Au collisions at $\sqrt{s_{_{\rm NN}}}$ = 200 GeV.  The left-hand
side of \Fref{fig:4_v2_ebye} shows the results for the average
midrapidity elliptic flow obtained from the event-by-event analysis
together with the results from both the hit-based and track-based 
analyses. The error bars represent statistical errors and the shaded
bands the 90\% C.L. systematic uncertainities.  Confidence that
all three measurements are determining the average elliptic flow 
is increased through the observation that they agree within 
the systematic errors.  The right-hand side of \Fref{fig:4_v2_ebye} 
presents the new PHOBOS results for $v_2$ dynamical fluctuations 
together with the result obtained for fluctuations in the participant
eccentricity.  Systematics on the experimental measurement are improved
by quantifying the result as a ratio of 
$\sigma(v_2)/\langle v_2 \rangle$.  We observe large dynamical
fluctuations in elliptic flow with a magnitude in remarkable agreement
with calculations of participant eccentricity fluctuations.  This result
suggests that the initial state equilibrates very rapidly with a
collision eccentricity largely defined by each collision event's particular
distribution of participant nucleon interaction points, 
and this detailed ``snapshot'' of the two nuclei's overlap region 
is propagated by the subsequent hydrodynamic evolution of 
the produced matter.

\begin{figure}[t]
\centering
\includegraphics[width=0.45\textwidth]{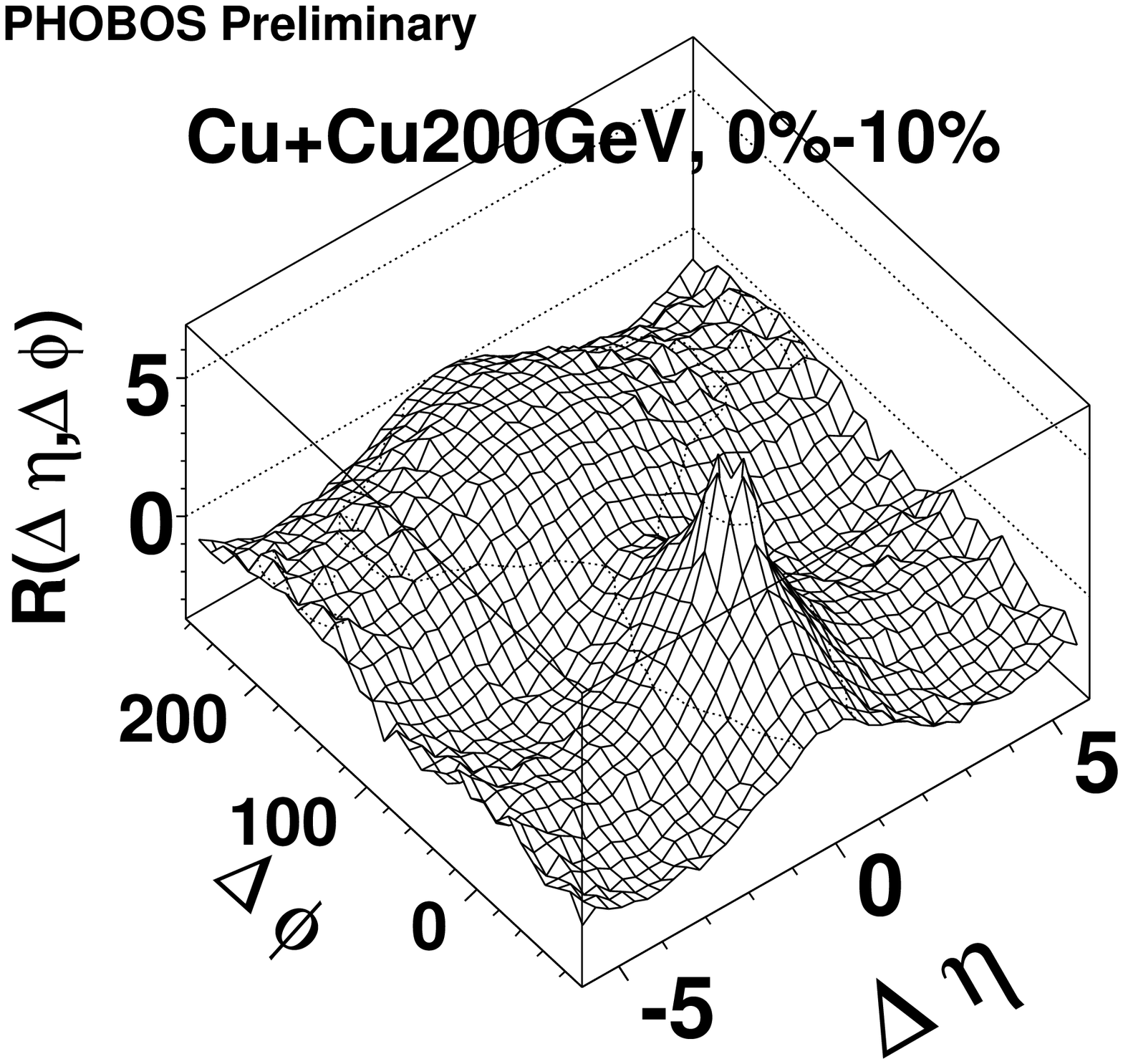}
\vspace{2pt}
%\hfill
\includegraphics[width=0.45\textwidth]{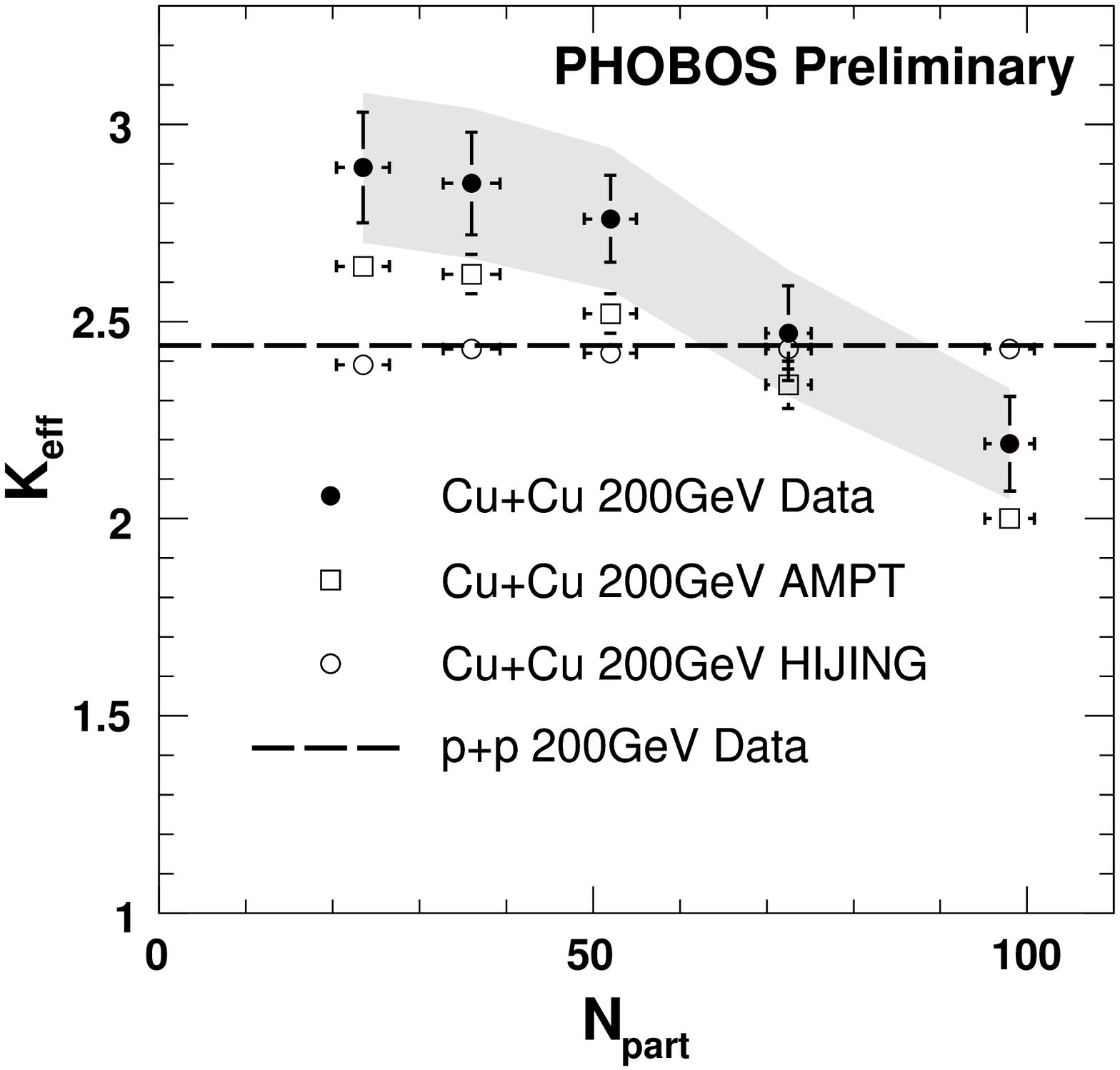}
\caption{Left: Two particle correlation in 
$\Delta \eta$ and $\Delta \phi$ between all charged particles for central
Cu+Cu collisions at $\sqrt{s_{_{\rm NN}}}$ = 200 GeV.  
Right: The extracted effective cluster size, $K_{\rm eff}$, in 
Cu+Cu and p+p data compared to HIJING and AMPT.
\label{fig:5_2partcorr}}
\end{figure}

PHOBOS has also made great progress in studying two-particle
correlations between charged particles.
We have recently published results on forward-backward
multiplicity correlations in non-overlapping bins of pseudorapidity for 
$\sqrt{s_{_{\rm NN}}}$ = 200 GeV Au+Au collisions \cite{PHOBOS:fb-corr}.
From this analysis we found that particles in Au+Au collisions are not 
produced independently, but instead appear to be produced in clusters.
In particular, we found significant short-range correlations at all
centralities, both as a function of $\eta$ and 
as a function of increasing pseudorapidity bin, $\Delta \eta$.
Here we report a completely new two-particle correlation
analysis that fully utilizes the extensive 
coverage in pseudorapidity ($| \eta | \leq 3.2$) and azimuthal 
angle ($\Delta \phi = 2 \pi$) afforded by the PHOBOS Octagon detector 
and its corresponding ability to measure essentially the 
full bulk of charged particle emission down to very low momentum.
At this point in time, we report the preliminary results of 
this measurement in both p+p and Cu+Cu collisions at 
$\sqrt{s_{_{\rm NN}}}$ = 200 GeV, where we emphasize that
no $p_T$ cut on any particle is imposed.
The correlation is calculated in the following way:
\[
R(\Delta \eta,\Delta \phi)=\left\langle (n-1) \left( 
\frac{F_n(\Delta \eta,\Delta \phi)}
{B_n(\Delta \eta,\Delta \phi)} - 1 \right) \right\rangle,
\]
where $F_n$ is the same-event two particle correlation between all
charged particles, $n$, and $B_n$ is the uncorrelated background.
An example of the results of 
this analysis for central Cu+Cu collisions is shown in the 
left-hand side of \Fref{fig:5_2partcorr}.  One can already observe
the $\Delta \phi$ correlations resulting from elliptic flow, something
not present in the corresponding p+p data.  Motivated by the interesting
results found in the forward-backward correlations analysis, we initially
have focused on studying the short-range correlations in pseudorapidity.
To accomplish this, we project onto $\Delta \eta$ to obtain one-dimensional
correlations in $\Delta \eta$, to which we perform fits in order to extract
the parameter $K_{\rm eff}$, the effective cluster size.  This result 
is presented in the right-hand side of \Fref{fig:5_2partcorr}.
The results clearly show that on average, even in heavy-ion collisions
such as Cu+Cu, particles tend to be produced in clusters with a size of 
$K_{\rm eff} = 2-3$, which is surprisingly similar to that seen in 
elementary p+p collisions.  In addition, there is a nontrivial 
dependence of $K_{\rm eff}$ on centrality, such that the value is
larger than seen in p+p collisions for peripheral Cu+Cu collisions 
and decreases with centrality to a value below that found in 
p+p for central collisions.  The observed variation of $K_{\rm eff}$
with centrality is not seen in HIJING, and although the centrality
dependence is qualitatively reproduced by AMPT, the magnitude is still
below that seen in the data. For more details see Ref.~\cite{Li:qm06}.

\begin{figure}[th]
\centering
\includegraphics[width=0.45\textwidth]{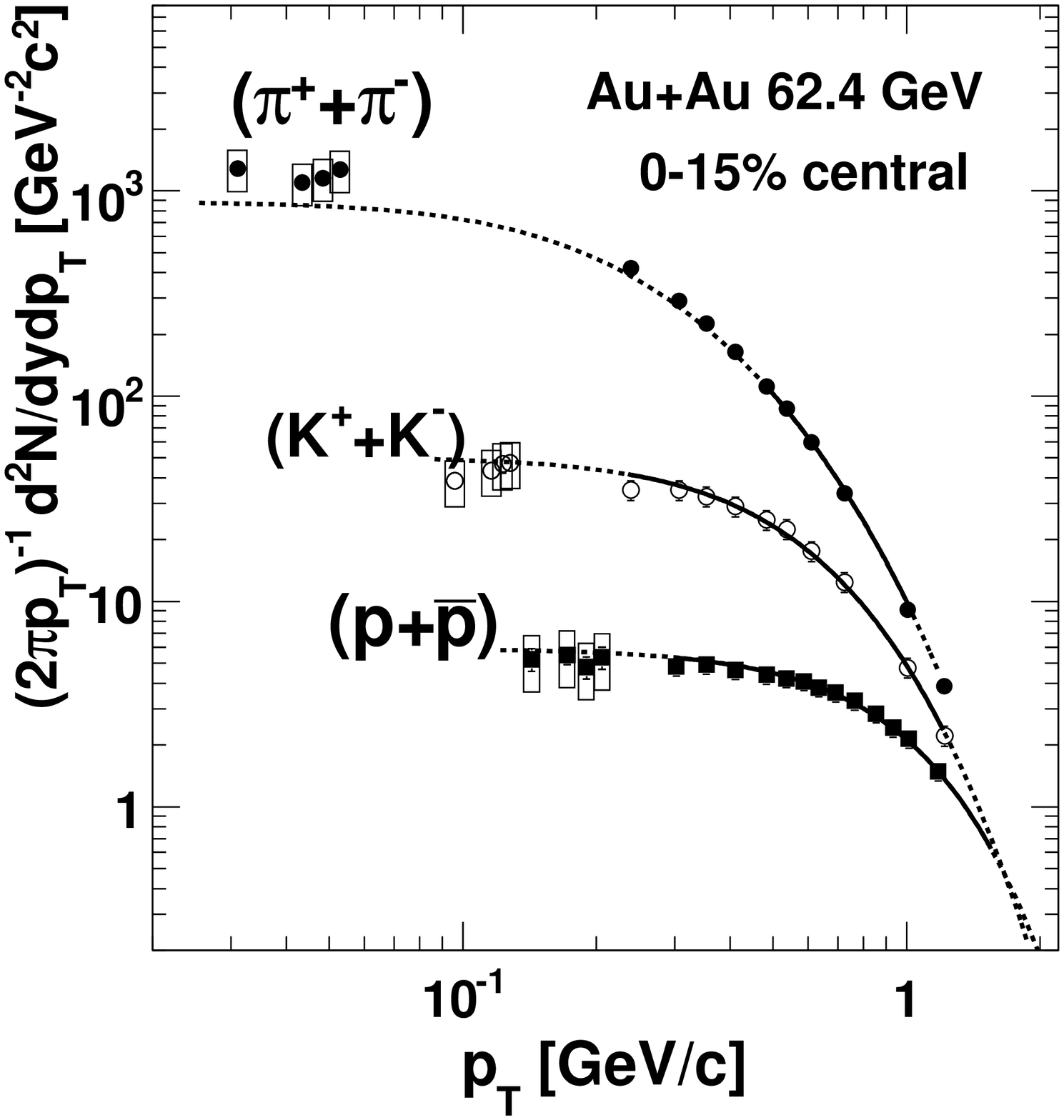}
\vspace{6pt}
%\hfill
\includegraphics[width=0.45\textwidth]{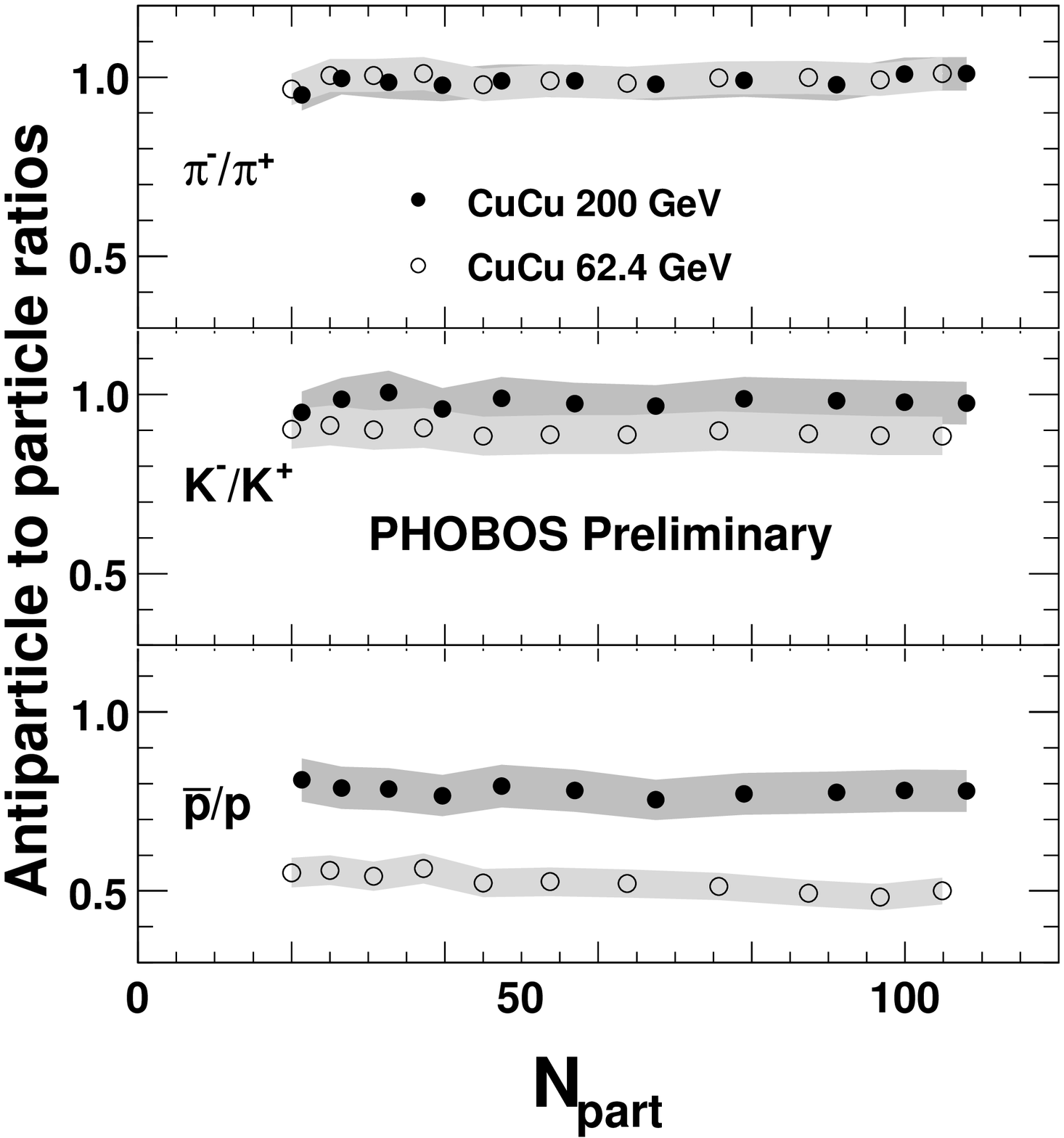}
\caption{Left: Identified particle spectra for Au+Au collisions 
at $\sqrt{s_{_{\rm NN}}}$ = 62.4 GeV. 
Right: Antiparticle to particle ratios for protons, kaons and 
pions in Cu+Cu collisions at $\sqrt{s_{_{\rm NN}}}$ = 
62.4 and 200 GeV. 
\label{fig:6_identpart}}
\end{figure}

PHOBOS has also continued its analysis of identified charged particles.
The left-hand side of \Fref{fig:6_identpart} gives one of the results
for identified hadron transverse momentum spectra 
in Au+Au collisions at $\sqrt{s_{_{\rm NN}}}$ = 62.4 GeV 
\cite{PHOBOS:spectra-auau}.  The unique coverage of PHOBOS at very 
low $p_T$ is evident, and the results of a blastwave fit to the higher
$p_T$ data, measured using both the Spectrometer and time-of-flight
detector, is found to be consistent with the combined yields of 
pions, kaons and protons at low $p_T$.  The right-hand side of 
\Fref{fig:6_identpart} shows new results on antiparticle to particle
ratios as a function of centrality measured in 62.4 and 200 GeV 
Cu+Cu \cite{Veres:qm06}.  
The PHOBOS measurement of antiparticle to particle ratios
is unique in that due to the two identical Spectrometer arms located on 
opposite sides of a dipole magnet we can, by simply reversing the
magnetic field systematically during data taking, extract four
independent measurements of identified particle ratios where all
effects of acceptance and efficiency cancel out in the final ratio.
The energy dependence of the proton and kaon particle ratios between
$\sqrt{s_{_{\rm NN}}}$ = 62.4 and 200 GeV is clearly evident
in \Fref{fig:6_identpart}, and we find in all cases that the 
ratios are only weakly dependent, if at all, on centrality.

\begin{figure}[htb]
\centering
\includegraphics[width=0.45\textwidth]{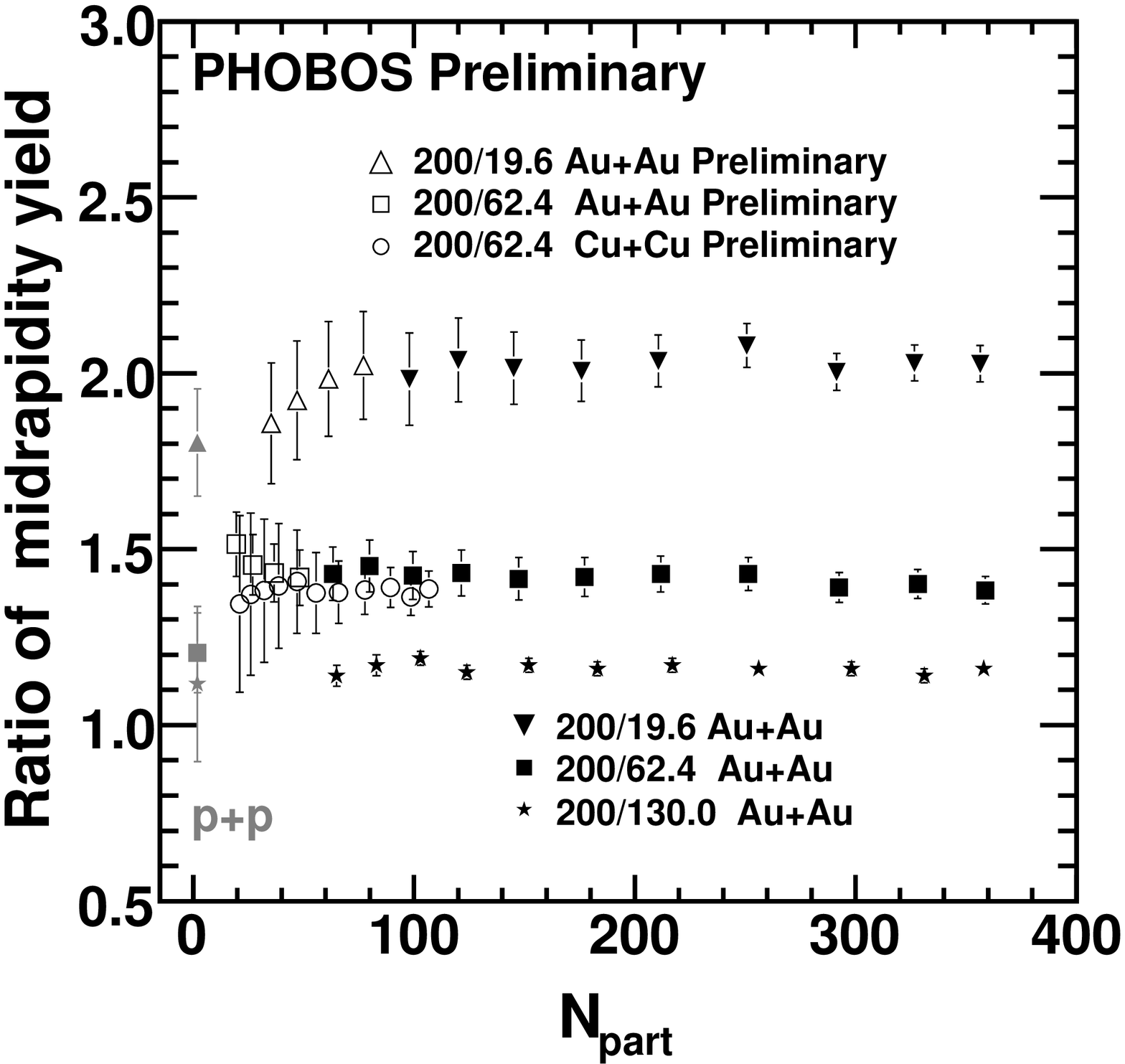}
\vspace{2pt}
%\hfill
\includegraphics[width=0.45\textwidth]{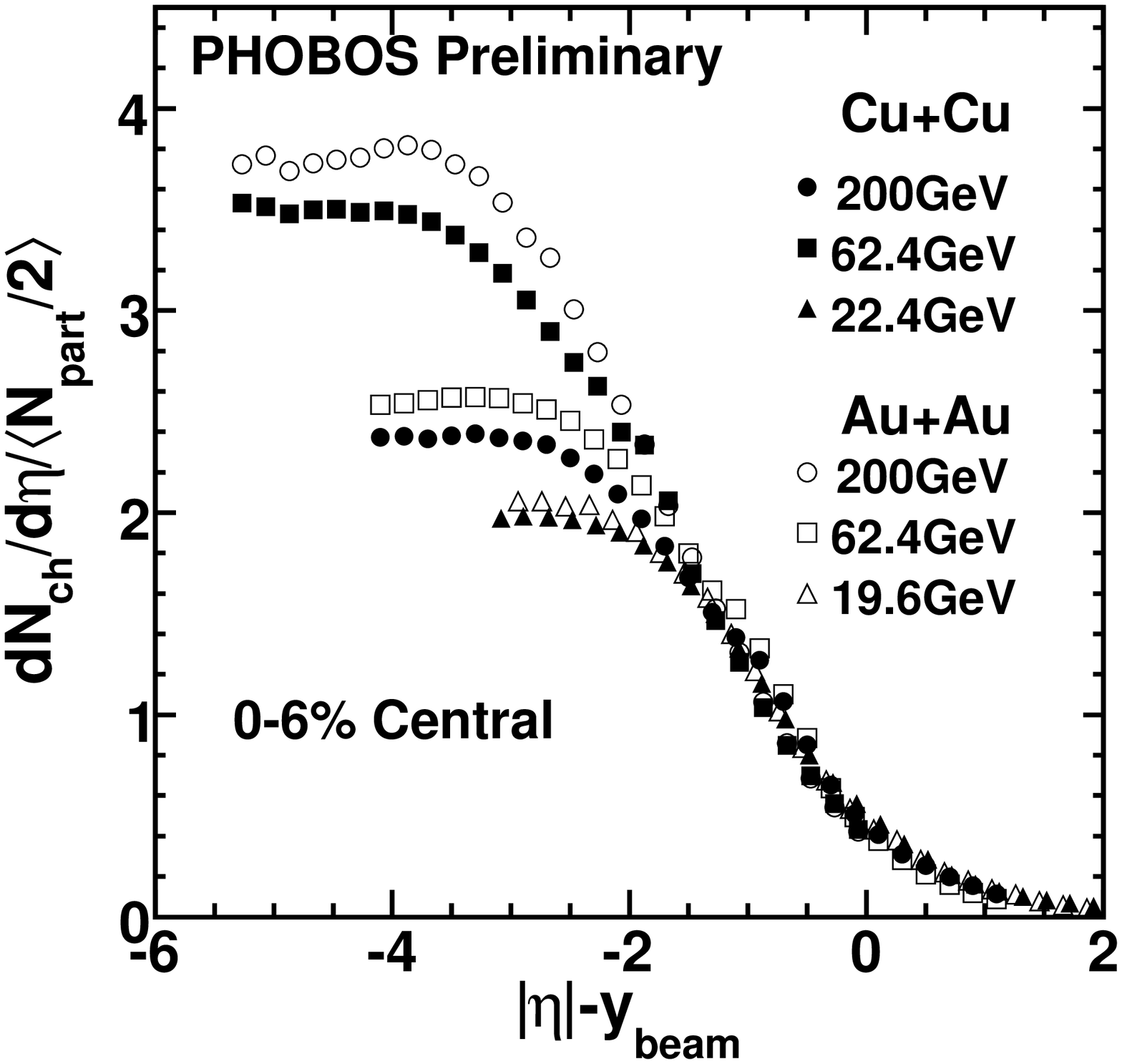}
\caption{Global charged particle production in Au+Au and Cu+Cu. 
Left: The ratio of midrapidity yields for different 
energies.  Right: The pseudorapidity 
dependence for central collisions effectively shifted into the rest frame 
of one of the nuclei. \label{fig:7_mult}}
\end{figure}

The latest results on global charged particle emission are shown in 
\Fref{fig:7_mult}.  As shown in the left-hand figure, we have extended 
the analysis for the midrapidity charged particle multiplicity in Au+Au
collisions to lower centrality, enabling a better overlap in $N_{\rm part}$ 
with the Cu+Cu system.  On the right-hand side of
\Fref{fig:7_mult}, we have also completed the analysis of 
$dN_{\rm ch}/d\eta$
for all Cu+Cu energies with the addition of the lowest energy data
at $\sqrt{s_{_{\rm NN}}}$ = 22.4 GeV.  The new results for even the 
lowest energy Cu+Cu data exhibit the same striking feature of extended 
longitudinal scaling \cite{PHOBOS:els} 
that appears to be a general feature of all 
heavy-ion results at RHIC energies.

The future PHOBOS physics program will continue to be based on the 
comprehensive dataset that already exists.  This data consists of 
Au+Au collisions at $\sqrt{s_{_{\rm NN}}}$ = 19.6, 62.4, 130 and
200 GeV, Cu+Cu collisions at $\sqrt{s_{_{\rm NN}}}$ = 22.4, 
62.4 and 200 GeV, d+Au collisions at $\sqrt{s_{_{\rm NN}}}$ = 200 GeV
and p+p collisions at $\sqrt{s_{_{\rm NN}}}$ = 200 and 410 GeV.
The p+p dataset will allow for a measurement 
of $dN_{\rm ch}/d\eta$ as a function of total multiplicity.  
On the heavy-ion side, we will extend the studies of dynamical 
flow fluctuations to 
other systems and energies.  The two-particle correlations study
between all particles will also be extended to Au+Au collisions.  
We are currently exploring the feasibility of performing two-particle
correlations studies using a high $p_T$ trigger particle, as 
measured in the Spectrometer, correlated with the bulk charged
particle production, as measured by the Octagon detector.  
We will continue to exploit the unique
PHOBOS capabilities for particle identification at very low $p_T$
by using the high-statistics Au+Au data set at $\sqrt{s_{_{\rm NN}}}$ = 
200 GeV to extract yields of pions, kaons and protons as a 
detailed function of centrality.  We plan to complete the PHOBOS
comprehensive study of antiparticle to particle ratios with measurements
in both Cu+Cu and Au+Au at $\sqrt{s_{_{\rm NN}}}$ = 62.4 and 200 GeV, 
as a detailed function of centrality and $p_T$. 
We also continue to actively pursue studies of $\phi$-meson production 
at very low $p_T$.  In short, although the active data-taking of PHOBOS
at RHIC has come to a close, there is still a great deal of
interesting and relevant physics to be explored and discovered.

\section*{Acknowledgments}

We acknowledge the generous support of the Collider-Accelerator Department
(including RHIC project personnel) and Chemistry Department at BNL.  We
thank Fermilab and CERN for help in silicon detector assembly.  We thank the
MIT School of Science and LNS for financial support.  

This work was partially supported by U.S. DOE grants 
DE-AC02-98CH10886,
DE-FG02-93ER40802, 
DE-FG02-94ER40818,  % MIT
DE-FG02-94ER40865,  % UIC
DE-FG02-99ER41099, and
DE-AC02-06CH11357, by U.S. 
NSF grants 9603486, % Phobos TOF 
0072204,            % Rochester until 6/03
and 0245011,        % Rochester starting 6/03
by Polish KBN grant 1-P03B-062-27(2004-2007),
by NSC of Taiwan Contract NSC 89-2112-M-008-024, and
by Hungarian OTKA grant (F 049823).

\end{document}